\documentstyle[aps,prl,multicol]{revtex}
\begin{document}
\renewcommand{\thefootnote}{\fnsymbol{footnote}}
\draft
\title{  Ladder operator for the one-dimensional Hubbard model}

\author {   Jon Links, Huan-Qiang Zhou, Ross H. McKenzie and
Mark D. Gould}

\address{Centre for Mathematical Physics,
School of Physical Sciences, \\
The University of Queensland, Brisbane 4072, Australia}

\maketitle


\begin{abstract}

The one-dimensional Hubbard model is integrable in the sense that it has
an infinite family of conserved currents. We explicitly construct a
ladder operator  which can be used 
to iteratively generate all of the conserved current
operators. This construction is different from that used for Lorentz
invariant systems such as the Heisenberg model. 
The Hubbard model is not Lorentz
invariant, due to the separation of spin and charge excitations.
The ladder operator is obtained by  a very general formalism which is
applicable to {\it any} model that can be derived from a solution of the
Yang-Baxter equation. 

\end{abstract}




\def\a{\alpha}
\def\b{\beta}
\def\e{\epsilon}
\def\g{\gamma}
\def\k{\kappa}
\def\l{\lambda}
\def\o{\omega}
\def\t{\theta}
\def\s{\sigma}
\def\D{\Delta}
\def\L{\Lambda}
\def\R{\cal{R}}
\def\h{\theta}  
\def\p{\partial}

\def\beq{\begin{equation}}
\def\eeq{\end{equation}}
\def\bea{\begin{eqnarray}}
\def\eea{\end{eqnarray}}
\def\ba{\begin{array}}
\def\ea{\end{array}}
\def\no{\nonumber}
\def\le{\langle}
\def\re{\rangle}
\def\lt{\left}
\def\rt{\right}
\def\dag{\dagger} 
\def\u{\uparrow} 
\def\d{\downarrow}
\def\nn{\nonumber}

\newcommand{\reff}[1]{eq.~(\ref{#1})}

\vskip.3in
\begin{multicols}{2}

The one-dimensional Hubbard model
has attracted considerable interest because
it is one of the few examples of a model
for strongly correlated electrons that is
exactly soluble\cite{ek94}. The fact that it describes
a doped Mott insulator and  exhibits
spin-charge separation (i.e., spin and charge
excitations are independent of one another)
has been argued to be relevant to understanding
the unusual metallic properties of high-temperature
superconductors\cite{anderson}.
Furthermore, the model has rich mathematical structure\cite{degkkk}:
it is solvable by both the co-ordinate Bethe ansatz\cite{lw} and
the algebraic Bethe ansatz\cite{mr}, it has a hidden $so(4)$
 symmetry\cite{yang}, and is integrable in the sense that
it has an infinite family of conserved 
currents\cite{sh}. The latter is 
a consequence of the fact that the model can be derived from a solution of 
the Yang-Baxter equation\cite{sw}.
Integrability is not just of mathematical
interest because it may have implications for
dissipationless transport\cite{zotos},
the coherence of interchain transport\cite{mila}, and whether the
energy level spacing follows
a Poisson distribution or 
the Gaussian orthogonal ensemble distribution characteristic of
quantum chaotic systems\cite{berry}.
Furthermore, integrability has been essential to recent
exact calculations of transport properties in mesoscopic
electronic devices\cite{ludwig}.

In this Letter we clarify the algebraic structure 
underlying the integrability  of the Hubbard 
model by using the Yang-Baxter equation
to explicitly construct a single operator $B$
(known as the ladder operator) which can be used in  a simple
recursion relation (equation (\ref{intro}) below) to generate
the whole family $\{ t^{(n)} \}_{n=0}^\infty$
of conserved current operators, i.e., operators
that commute with the Hamiltonian and one another.
This result is surprising in light of the  
lack of invariance in the model under the lattice
version of the Poincare group.

For continuum field theories in 1+1 dimension
the generators of the Poincare group are 
$B, \, P$ and $ H$, being the generators of Lorentz boosts 
and translations in space and time, respectively.
($P$ and $H$ are also the total momentum operator
and Hamiltonian, respectively).
They obey the closed algebra 
\begin{equation}
[B,\,H]=P,~~[B,\,P]=H,~~[H,\,P]=0.
\end{equation}
It is extraordinary that a wide range of integrable
lattice models (including the 
 Heisenberg\cite{thacker,sogo},
 Calogero, Toda,\cite{sogo} and supersymmetric $t-J$\cite{essler}
models) are invariant under 
a generalization of the Poincare group
involving the entire infinite set
$\{ t^{(n)} \}_{n=0}^\infty$ of conserved currents.
They satisfy the algebra
\beq [B,\,t^{(n)}]=t^{(n+1)},~~ \, \, \, \,
\, \, [t^{(n)},t^{(m)}] = 0
\label{intro} \eeq   
where $t^{(0)}$ and $t^{(1)}$ are the momentum operator
$P$ and Hamiltonian $H$, respectively.
Here the boost operator 
acts as a ladder operator on the infinite
sequence of conserved operators.
For the $XXZ$ model the boost operator can 
be identified with an algebraic element of
a lattice Virasoro algebra\cite{t}. This turns out
to be of great practical significance because
it permits the use of vertex operators for the
determination of the energy spectrum and calculation of correlation
functions\cite{jm94}.

A crucial property in the manifestation of Lorentz invariance in the
above models is the fact that
the $R$ matrix which is a solution of the Yang-Baxter equation
(equation (\ref{ybe}) below)
has the difference property, $R(u,v)= R(u-v)$
for the spectral parameters $u$ and $v$.
This is because  the spectral
parameter plays the role of rapidity variable. A uniform shift in both
rapidity variables, corresponding to a change in the Lorentz frame,
leaves the $R$-matrix invariant. In this sense those solutions with
the difference property are invariant under a lattice version
of the Poincare group\cite{thacker}.

In contrast, the Hubbard model
is not Lorentz invariant \cite{frahmkorepin} since it exhibits gapless
excitations with different velocities. It is for this reason that spin
and charge separate. As a result, the $R$ matrix associated with the
Hubbard model does not have the difference property
and so its integrability 
is not as well understood. Although Lieb and Wu\cite{lw}
gave a co-ordinate Bethe ansatz solution
in 1968, it was not until 1986 that Shastry
demonstrated the existence of an infinite family
of conserved currents. This involved constructing
a two-dimensional model in classical statistical mechanics
with a transfer matrix that commuted with the Hubbard
Hamiltonian\cite{sh}. 
It was achieved by mapping the model onto a
pair of coupled spin chains using the Jordan-Wigner transformation.
Although Shastry conjectured that the
$R$-matrix satisfies the Yang-Baxter equation, it was some time
before a convincing proof was available \cite{sw}. 
Furthermore, the  use of the algebraic Bethe ansatz method 
to reproduce the solution of Lieb and Wu has only recently been
achieved \cite{mr}. A significant consequence of this algebraic
development is that it facilitates the use of the  quantum transfer
matrix method for the analysis of the thermodynamic properties at
finite temperature \cite{degkkk,jks}. 

Grabowski and Mathieu \cite{gm} claimed that there is no
``matrix'' ladder
operator satisfying (\ref{intro}) for the model, motivating them to construct
the first seven conserved currents by ``brute force methods''.
We now show how for {\it any} model derived from a solution of the
Yang-Baxter equation  there is a one parameter family
of ladder operators $B(v)$ such that the conserved currents
satisfy (\ref{intro}).   
In cases where a solution to the Yang-Baxter equation
has the difference
property, the conserved currents  have no dependence on the spectral
parameter $v$. 
In this instance the construction for the
ladder operator occurs as a particular case of the more general method we
describe below. 

As an application of our general result we then consider a one-parameter
of Hamiltonians which include the Hubbard model as a special
case ($v=0$). Our approach has the further appeal
that we work directly with the fermion operators
of the model (rather than a two-dimensional statistical
mechanics model) and that the $so(4)$ invariance of
the model is manifest throughout.

The Yang-Baxter equation (or star-triangle relation)
is central to exactly soluble models because it
 a sufficient condition
for the validity of the Bethe ansatz \cite{b,rc}.
The corresponding equations for $1 + 1$ dimensional 
quantum field theories are also known as the
factorisation equations because they imply
that all possible decompositions of the
N-particle scattering (S) matrix give the same
result as a product of two-particle S matrices\cite{zz}.
Consider a lattice model defined on $L$ sites, each
of which has a Hilbert space $V$.
The matrix $R(u,v)$ acts on the tensor product space
$V \otimes V$  and satisfies
the Yang-Baxter equation \cite{b,rc} 
\bea 
&&R _{12}(u,w)  R _{13}(u,v)  R _{23}(w,v) \nn \\
&&~=
R _{23}(w,v)  R _{13}(u,v)  R _{12}(u,w).
\label{ybe}  \eea 
where the subscripts refer to the embedding of $R(u,v)$ on the 3-fold
space $V\otimes V\otimes V$. 
From a solution to the 
this        equation we define a transfer matrix 
$$t(u,v)={\rm tr}_0\left(R_{0L}(u,v)...R_{02}(u,v)R_{01}(u,v)\right)$$
where tr denotes the trace over $V$ (when $V$ is a superspace we 
use the supertrace, i.e., the trace over
the bosonic states minus the trace over the fermionic states).
It follows from the Yang-Baxter equation (\ref{ybe})  that 
\beq [t(u,v),\,t(w,v)]=0 \label{ctm} \eeq
for all values of the parameters $u$ and $w$.  

An assumed  feature of  the  $R$ matrix is the regularity property, i.e., 
$R(u,u) = P$, with $P$ being the permutation operator.
(A phase of (-1) is gained whenever two fermionic states are
interchanged.)
Using this property, the Hamiltonian is defined to be 
\beq  
H(v)=-T^{-1}.\left.\frac{\partial t(u,v)}{\partial u}\right|_{u=v}
\label{localham} \eeq  
where  $T\equiv t(u,u)=P_{1L}...P_{13}P_{12}$ is the translation operator.
This yields
\beq H(v)=\sum_{j=1}^L h_{j(j+1)}(v)\label{ham} \eeq
with the local Hamiltonian given by 
$$h(v)=-P \left.\frac{\partial R(u,v)}{\partial u}\right|_{u=v}.$$
Above and throughout periodic boundary conditions
are imposed. 
For later use, it is convenient to consider the series expansion for the
$R$-matrix 
\bea  
R(u,v)&=&P(I+(v-u)h(v)+1/2(v-u)^2f(v)\nn \\
&&~+1/6(v-u)^3g(v)+...) 
\label{expan} \eea 

Expressing the logarithm of the transfer matrix in a power series
expansion
\beq
\ln t(u,v)=\sum_{n=0}^{\infty} \frac{(u-v)^n}{n!} t^{(n)}(v), 
\label{log} \eeq 
it is apparent,
in view of (\ref{ctm}), that
$$[t^{(n)}(v),\,t^{(m)}(v)]=0,~~~\forall\,m,n$$
and in particular
$$[H(v),\,t^{(n)}(v)]=0.$$
Consequently, the Hamiltonian $H(v)$ is integrable since the set of
operators $\{t^{(n)}(v)\}$ provide a set of conservation laws for the
system.
Note that in the case where the $R$-matrix does not have
the difference property, we have the generic feature that the
Hamiltonian and higher conserved charges
will always have non-trivial dependence on the variable
$v$, as can be seen from (\ref{log}).

We now construct the parameter-dependent ladder operator $B(v)$.
Differentiating the Yang-Baxter equation (\ref{ybe})
with respect to $w$, then setting $w=v$
and premultiplying by the permutation
 operator $P_{j(j+1)}$ yields an analogue of
the Sutherland equation \cite{su}
\bea 
&&[h_{j(j+1)}(v),\,R_{0(j+1)}(u,v)R_{0j}(u,v)] \nonumber \\
&&~~= 
R_{0(j+1)}(u,v) \frac{\partial R_{0j}(u,v)}{\partial v} 
-\frac{\partial R_{0(j+1)}(u,v)}
{\partial v}R_{0j}(u,v). \label{se} \eea 
An immediate consequence of (\ref{se}) is that 
$$[H(v),\,t(u,v)]=0$$ 
which also follows from (\ref{ctm}). 

The ladder operator $B(v)$
is defined in terms of the local Hamiltonians $h_j\equiv h_{j(j+1)}(v)$ 
through the relation
\beq
B(v)= -\sum_{j=1}^L {\bf j} h_j  +\frac{\p }{\p v} \label{bo} 
\eeq
where ${\bf j}$ are the elements of the
integers modulo $L$. A consequence of the generalized Sutherland
relation (\ref{se}) is now
\beq
[B(v),\, t(u,v)] =  0. 
\label{boost} \eeq
The relation
(\ref{boost}) permits us to deduce the  recurrence relation
(\ref{intro})
from (\ref{log}).

The definition of the ladder operator here is
different from the cases considered in 
\cite{thacker,sogo,essler,t} by the inclusion of 
the differential operator. This term is not required for those cases
with the difference property since it is apparent from (\ref{log}) 
that the conserved
currents $t^{(n)}$ have no dependence on $v$ and hence (\ref{intro})
still holds.  
However, in this instance (\ref{boost}) becomes $$[B,\,t(u-v)]=\frac{\p
t(u-v)}{\p u}$$ which  can be integrated to 
$$ t(u+\lambda)=\exp(\lambda B)t(u)\exp(-\lambda B).$$ 
The
parameter $u$ characterizes the Lorentz frame for the transfer matrix
$t(u)$, showing that $B$ is the
generator of Lorentz boosts in this context\cite{thacker}. This is clearly 
not the case for the
Hubbard model where Lorentz invariance is not present.

From (\ref{intro}) the operators $t^{(n)}(v)$ may be
calculated iteratively. We find the following expressions for the
leading terms
\bea 
t^{(0)}&=&\ln T, \no \\
t^{(1)}&=&-H, \no \\
t^{(2)}&=&\sum_{ij}\,{\bf j}[h_j,\,h_i]-H' \nn \\
&=&\sum_j\,{\bf j}[h_j,\,h_{j+1}+h_{j-1}]-H' \nn \\
&=&\sum_j\,{\bf j}[h_j,\,h_{j+1}]+\sum_j({\bf j+1})[h_{j+1},\,h_{j}]-H' \nn
\\
&=&-\sum_{j}[h_{j},\,h_{j+1}]-H' 
\nn \eea
where the prime denotes a derivative with respect to $v$.
The computation of $t^{(3)}$ can be simplified
by   invoking the generalized Reshetikhin
condition (cf. \cite{rc}) 
\bea &&[h_{12}+h_{23},\,[h_{12},\,h_{23}]] \nn \\
&&~~+[h_{12},h'_{12}]+[h_{12},\,h'_{23}]+[h_{23},h'_{12}]\nn \\
&&~~~~= x_{23}-x_{12} \label{rc} \eea
with the two site operator given by
$$x=2h^3+g-3hf+2[h',\,h]-h''. $$
The Reshetikhin relation is obtained by applying $\partial^3/\partial
u^2\partial w$ to (\ref{ybe}) and using (\ref{expan}). 
Omitting the details, this yields the result
\bea
t^{(3)}&=& 
cH+2\sum_j[h_j,\, [h_{j-2},\,h_{j-1}]+h'_{j-1}]
\nn \\
&&~~+\sum_j[h_{j},\,[h_{j-1},\,h_{j}]-h'_{j+1}] \nn \\
&&~~~+\sum_j (h_j^3-g_j-h_j.h_j'-2h_j'.h_j). 
\nn \eea
Above, $c$ is a constant determined by the normalization of $R(u,v)$. 
It can always be chosen to be zero. 
Before turning to the particular case of the Hubbard
model we stress that the above construction of the ladder 
operator is valid for any solution of the Yang-Baxter equation
(\ref{ybe}).

Consider the four dimensional
local Hilbert space $V$ spanned by 
the  states 
$$\left|0\right>,~\left|\u\right>,~\left|\d\right>,~\left|\u\d\right>.$$
Introduce the spaces $W(\s)$ with basis
$\{\left|0\right>,~\left|\s\right>\},~\s=\u\d$ so that  
$V\equiv~W(\u)\times
W(\d)$. For each tensor space $W(\s)\otimes W(\s)$ there is a 
 solution of the Yang-Baxter equation (with difference property) given by 
\bea
{\R}_{ij}^{\s}(u-v)&=&\cos (u-v)(1-n_{i\s}-n_{j\s})
\nn \\
&&~+\sin (u-v)(n_{i\s}
+n_{j\s}-2n_{i\s}n_{j\s}) \nn \\
&&~+c^{\dag}_{i\s}c_{j\s}+c^{\dag}_{j\s}c_{i\s}. \nn \eea
Here  $ c^{\dagger}_{j\s}$ and $ c_{j\s}$ are the creation and
annihilation
operators with spin $ \s (= \uparrow,\downarrow) $ at site $j$ and
$n_{j\s} = c^{\dagger}_{j\s}c_{j\s} $ is the density operator.
The associated Hamiltonian obtained through (\ref{localham}) is that for
free fermions.

It has recently been shown\cite{usw} that the following $R$-matrix  
is also a solution of the Yang-Baxter equation acting on $V\otimes V$,  
\bea 
R_{ij}(u,v)&=&{\R}_{ij}^{\u}(u-v){\R}_{ij}^{\d}(u-v)\nn \\
&-&\frac{\cos(u-v)}
{\cos(u+v)}\tanh\left(\h(u)-\h(v)\right) \nn \\
&\times &{\R}_{ij}^{\u}(u+v){\R}^{\d}_{ij}(u+v)(1-2n_{i\u})(1-2n_{i\d})
\label{rm} \eea
with 
$\h(u)$ defined through
the relation
$$\sinh 2\h(u)=\frac U4 \sin 2u.$$   

An important consequence of (\ref{ybe}) is that it allows for the
construction of a generalized Hubbard model (with spectral parameter 
dependence) as noted in \cite{usw}. 
The identification of this generalized model is paramount in the
construction of the laader operator. 
Explicitly, the local Hamiltonians read
\bea
h_{ij}(v)&=&-\sum_{\s=\u\d}\left(c^{\dag}_{i\s}c_{j\s}+c^{\dag}_{j\s}c_{i\s} 
\right) \nn \\
&&~~+\frac{U}{4\cosh 2\theta(v)}\Gamma_{ij\u}(v)\Gamma_{ij\d}(v)
\label{locham}
  \eea
where
\bea \Gamma_{ij\s}(v)&=&\cos^2v(1-2n_{i\s})-\sin^2v(1-2n_{j\s})\nn \\
&&~~+\sin 2v(
c^{\dag}_{i\s}c_{j\s}-c^{\dag}_{j\s}c_{i\s}). \label{genhub} \eea 
It is clear that (\ref{ham}) with
(\ref{locham}) reduces to the usual Hubbard model  when
$v=0$. 
In particular, we find 
\bea 
h'_{ij}(0)&=& U/2\left[(1-2n_{i\u})(c^{\dag}_{i\d}c_{j\d}-c^{\dag}_{j\d}c_{i\d})
\right. \nn \\
&&~~+\left.(1-2n_{i\d})(c^{\dag}_{i\u}c_{j\u}-c^{\dag}_{j\u}c_{i\u})
\right] \label{hubprime} 
\eea     
which plays an important role in the explicit construction of the higher
conserved operators discussed previously. It is important to note that
there is no way of determining $h'(0)$ directly from
the usual Hubbard model.

Substituting $g$ (which is obtained through 
(\ref{expan}) and (\ref{rm})) and eqs. (\ref{genhub},\ref{hubprime}), all  
evaluated at $v=0$, into the above expressions for $t^{(2)}$ and $t^{(3)}$, we
recover the expressions found previously
for the first\cite{sh}
and second\cite{gm,zjt} non-trivial conserved currents 
(modulo a constant term and multiple of $H$). 
A well known feature of the Hubbard model is that
all the integrals of motion, except for the
translation operator, are invariant with respect to the $so(4)$ Lie
algebra 
when the lattice length is even\cite{usw}.
Significantly, the ladder operator (\ref{bo})
in the case of the Hubbard model 
is also $so(4)$ invariant for an even length lattice.  
%

To conclude, a systematic method for obtaining the conserved currents
in the Hubbard model has been described which employs the use of a ladder
operator. We emphasize, however, that the construction presented here is
entirely general and may be applied to any Yang-Baxter integrable
system. 

The authors thank the Australian Research Council for financial support.

\end{multicols}

\end{document}